\documentclass[twocolumn]{aastex631}
\usepackage{amsmath}

\newcommand{\be}{\begin{equation}}
\newcommand{\ee}{\end{equation}}
\defcitealias{LC}{TSO}

\begin{document}

\title{Strong Scatterings Invalidate Proposed Models of Enhanced TDE Rates in Post-Starburst Galaxies}

%Strong scatterings challenge stellar properties scenarios for the
%post-starburst preference of Tidal Disruption Events  }
\correspondingauthor{Odelia Teboul}
\email{odelia.teboul@weizmann.ac.il}

\author[0009-0006-1177-7466]{Odelia Teboul}
\affil{Technion – Israel Institute of Technology, Haifa, 3200002, Israel}

\author{Hagai B. Perets}
\affil{Technion – Israel Institute of Technology, Haifa, 3200002, Israel}
\affil{Astrophysics Research Center of the Open University (ARCO), \\The Open University of Israel, P.O. Box 808, Raa'nana 4353701, Israel
}

%\begin{document}

%\maketitle

\begin{abstract}
%The overrepresentation of tidal disruption events (TDEs) in post-starburst galaxies suggests a possible link between galaxy evolution and the rate of TDEs. To explain this preference, previous studies have explored the effects of ultra-steep stellar density profiles and radial velocity anisotropy on TDE rates in nuclear star clusters. In this study, we extend these previous works by including the effects of strong scattering, a process that can eject stars from the loss cone. We examine the combined effects of strong scattering, velocity anisotropy, and ultra-steep density profiles on TDE rates, considering a range of realistic stellar initial mass functions. Our results indicate that the inclusion of strong scattering significantly reduces or eliminates the TDE rate enhancements predicted by ultra-steep density profiles or radial velocity anisotropy alone. This finding challenges the proposed explanations for the post-starburst preference of TDEs and suggests that additional factors may be at play. We discuss alternative mechanisms, such as binary supermassive black holes and secular effects in eccentric stellar disks, that could potentially explain the observed preference for TDEs in post-starburst galaxies.

Stars wandering too close to supermassive black holes (SMBHs) can be ripped apart by the tidal forces of the black hole. Recent optical surveys have revealed that E+A galaxies are overrepresented by a factor $\sim $ 30, while green galaxies are overrepresented in both optical and infrared surveys. Different stellar models have been proposed to explain this Tidal Disruption Event (TDE) preference: ultra-steep stellar densities in the nuclear cluster, radial velocity anisotropies, and top-heavy Initial Mass Function (IMF). Here we explore these hypotheses in the framework of our revised loss cone theory that accounts for both weak and strong scattering, i.e., a scattering strong enough to eject a star from the nuclear cluster. We find that, when accounting for weak and strong scatterings, both ultra-steep densities and radial velocity anisotropies fail to explain the post-starburst preference of TDEs except when considering a high anisotropy factor together with a high SMBH mass and a shallow density profile of stellar mass black holes $\gamma_{\rm bh} =7/4$. Our findings hold when combining either model with top-heavy IMFs. Hence, new models to explain the post-starburst preference of TDEs are needed. 

%except for a high SMBH together with high anisotropy factor and a shallow density profile of stellar mass black holes $\gamma_{\rm bh} =7/4$ could . Our findings hold when combining either model with top-heavy IMFs. 
%Together we find that rate enhancements in E+A/green-valley galaxies pose a challenge to current models of TDE production.      
\end{abstract}

\section{Introduction}
\label{sec:Intro}
An unlucky star wandering too close to an SMBH can be torn apart by the SMBH’s tidal forces. Roughly half of the gaseous debris from the disrupted star falls back onto the SMBH powering an extremely luminous, multi-wavelength electromagnetic flare that will typically outshine the entire host galaxy for a few months \citep{Rees88}. TDEs were first theoretically predicted \citep{Hills75, Lacy, Rees88, EvansKochanek89}
and only detected at the advent of X-ray all-sky surveys by ROSAT
only later, with the advent of X-ray all-sky surveys by ROSAT \citep{Bade}. 

%, were nuclear transients have been observed and suggested to be TDEs (but see \cite{Saxton+18,Zab+21} for alternative possibilities for the origins of these transients).

%The number of observed TDEs has significantly increased thanks to the advent of all-sky surveys. 
The sample of observed TDEs has been growing at an increasing rate and now spans all wavelengths, from the radio to gamma-rays. Optical surveys revealed a surprising over-representation of these events in post-starburst galaxies \citep{Arcavi+14, French+16, French+17, Law, Graur, French20, Hammerstein21}. This overrepresentation was initially estimated to be $\sim 100-190$ in E+A galaxies, galaxies that makeup $\sim 0.2 \%$  of low-redshift galaxies, and whose recent starburst created $ >3 \%$ of their current stellar mass over $25–200$ Myr \citep{French+16}. Quiescent Balmer-strong galaxies which make up $\sim 2\%$ of
local galaxies, and which formed $>0.1 \%$ of their current
stellar mass over $25$ Myr $ - 1$ Gyr exhibited a smaller boost of $\sim 30$ \citep{French+16}. \cite{Law} showed that accounting for selection effects could reduce the TDE boost in E+A galaxies to a factor $\sim 25-48$, while the latest estimations suggest that E+A galaxies are overrepresented by a factor of $ \sim 22-29 $ \citep{Hammerstein21}. A less significant preference for green valley galaxies was also observed in optical surveys \citep{Hammerstein22, Yao+23} as well as in Infrared (IR) survey \citep{IR}. 

%of a dozen TDEs showed that IR-selected TDEs do not seem to have a preference for E+A galaxies, but still harbour a preference for green galaxies \citep{IR}.

Several hypotheses have been proposed to explain this puzzling preference for post-starburst galaxies. 
Assuming that post-starburst galaxies stem from galaxy mergers, \cite{Arcavi+14} proposed that the post-starburst preference could be triggered by SMBH binaries. \cite{Madigan+18} found that an eccentric nuclear disk could significantly enhance TDE rates and hence explain the preference if eccentric nuclear disks form during galaxy mergers. The presence of an AGN disk was also found to increase TDE rates by a factor of $\sim 10$ (e.g., \cite{Kennedy, Kaur}; however \cite{Wang24} found larger enhancements). The interaction of stars with massive perturbers and/or nuclear spiral arms could also slightly increase TDE rates, by up to a factor of two \citep{Perets+07,Hamers:2017}. 
%We briefly discuss those hypotheses in the discussion.

%However, the not long-lasting

%Arcavi et al. (2014) proposed that, as many starbursts are triggered by galaxy mergers, the PSP may reflect a population of SMBH binaries (SMBHBs) which, as they harden, pass briefly through a stage where TDE rates are in- creased by many orders of magnitude. Stone & Metzger (2016) suggested that if the starburst is preferentially concentrated in the galactic nucleus, then a strong stellar overdensity may be formed, which would increase TDE rates by decreasing the two-body re- laxation timescale.  Other possibilities exist as well: rates could be enhanced due to non-conservation of or- bital angular momentum in a triaxial potential created by starburst

Other proposed explanations invoke Nuclear Star Cluster (NSC) star characteristics: ultra-steep stellar densities with  $\rho \propto r^{-\gamma_\star} $,  $\gamma_\star \geq 9/4  $ \citep{Stone+18}, radial velocity anisotropies  \citep{Stone+18} and a complete stellar function with a top-heavy IMF \citep{Bortolas}. Both ultra-steep stellar densities and radial velocity anisotropies were found to increase TDE rates by factors up to a few hundreds. 

%\cite{Stone+18} found that
%could induce, in the most favorable cases, an enhancement of a factor $\sim 100$. 
%making these hypotheses appalling to explain the lasting preference of TDEs for post-starburst galaxies.  

%These explanations can easily be separated on two families of hypotheses. The first family englobes disc/SMBH characteristics: , while the second family invokes Nuclear Star Cluster (NSC) star characteristics: ultra-steep stellar densities \cite{}, radial velocity anisotropies \cite{} and top-heavy IMFs \cite{Bortolas}. 

%However, the observed per-galaxy TDE rate has been constrained to a range $\dot{N}_{\rm TDE} \sim 10^{-5} - 10^{-4}$ yr$^{-1}$~ gal$^{-1} $ by all-sky surveys \citep{Holoien16, vanVelzen18}. Moreover, these optical surveys revealed a surprising over-representation of these events in post-starburst (e.g., ``E+A'') galaxies \citep{Arcavi+14, French+16,French+17, Law, Graur, French20, Hammerstein21}. The latest estimations suggest that E + A galaxies are overrepresented by a factor of $ \sim 22 $ \citep{Hammerstein21}. 

In addition to this puzzling post-starburst preference, optical surveys observed TDE rates were constrained to a range $\dot{N}_{\rm TDE} \sim 10^{-5} - 10^{-4}$ yr$^{-1}$~ gal$^{-1} $ \citep{Holoien16, vanVelzen18, Yao+23}.
% further difference non E+A galaxies, empirically 
However, theoretically predicted rates computed with classical loss cone theory were typically estimated in the range $\dot{N} \sim 10^{-4} - 10^{-3}$ yr$^{-1}$~ gal$^{-1}$ (e.g., \cite{WangMerritt04, StoneMetzger16}). This discrepancy between observed and theoretically predicted rates worsens in non-E+A galaxies, as the post-starburst preference of TDEs observed in several surveys further reduces the observed rates in non-E+A galaxies.

%TDE rates are traditionally estimated from classical loss cone theory, which focuses on the cumulative effect of many weak scatterings. These scatterings are modeled as local and uncorrelated, and lead to effective diffusion coefficients describing the drift and diffusion of stellar populations through phase space \cite[see][for reviews]{Merritt13, Stone+20}. Strong, or small impact parameter scatterings are generally neglected, as they are largely outnumbered by weak scatterings.

%A star wandering too close to an SMBH can be ripped apart by the SMBH’s tidal forces. Roughly half of the gaseous debris from the disrupted star falls back onto the SMBH, eventually circularizing into an accretion disk and powering an extremely luminous, multi-wavelength electromagnetic flare that will typically outshine the entire host galaxy for a few months \citep{Rees88}.
Classical loss cone theory calculations are carried out by resolving the Fokker-Planck equation that focuses on the cumulative effect of many weak scatterings. Strong or small impact parameter scatterings are generally neglected, as they are largely outnumbered by weak scatterings. However, most stars that become TDEs come from within the radius of influence, {\it the densest environments of the Universe}. In such environments, close encounters rare elsewhere can become non-negligible. 

Hence, we proposed a revised loss cone theory taking into account both weak interactions and other close encounters: strong scattering, tidal captures, and direct collisions (\cite{LC}, hereafter \citeauthor{LC2}). 
% TDE rate calculations were carried out with classical loss cone theory that takes into account two-body weak interactions between stars.  
 We found that, at the radius of influence, the dominant mechanism is strong scattering, i.e. a scattering strong enough to eject the test star from the distribution. 
 %Then, we derived analytical solutions to the Fokker-Planck equation with strong scattering for the different black hole slopes.  
 We showed that, depending on the black hole density slope $\gamma_{\rm bh}$, {\it TDE rates are reduced by up to an order of magnitude}, reconciling them with observed TDE rates (e.g \cite{Yao+23}). 
 
 %In this paper we will revise the standard loss cone theory to take into account not simply weak two-body scatterings, but also destructive processes such as strong scattering ejections, collisions,  and tidal captures, with the ultimate goal of determining under which conditions close encounters can {\it shield the loss cone}. 
 
The black hole density slope depends on the segregation mode. Indeed,  in a system composed of both stars and heavier objects, the heavier objects are expected to segregate towards the center of the galactic nucleus and settle on a steeper cusp while the light objects will have a weaker cusp $\gamma_\star \approx 1.3 -1.5 $. In the weak segregation limit, the black hole slope assumes $ \gamma_{\rm bh} = 7/4 - 2$  \citep{BW77, PretoAmaroSeoane10, Amaro, Broggi} whereas in the strong segregation limit, the heavy objects have been predicted to settle to even steeper power-law slopes of $\gamma_{\rm bh} = 2 - 11/4  $ (\cite{AlexanderHopman09}; see also \cite{Zhangpau,Aha+16}). 
%Moreover for an  adiabatically growing MBH, the stellar slope may assume steeper values  of $\gamma \gtrsim 2$ \citep{Young}.

Here we investigate the proposed explanations invoking NSC star characteristics for the post-starburst preference of TDEs in the framework of our revised loss cone theory that takes into account both weak and strong scattering. 
%We will consider time-dependant Motivated by the importance of    

We present classical loss cone theory in Section \ref{sec:LC},   briefly summarize the key results of strong scattering in Section \ref{sec:2.1}, and  present our modified Fokker-Planck equation with strong scattering in  Section \ref{sec:2.2}. In Section \ref{sec:anisotropies} we explore the effect of radial velocity anisotropies with and without strong scattering, while in Section \ref{sec:overdensities} we investigate the impact of ultra-steep stellar densities. In Section \ref{sec:IMF}, we combine the effect of different Present Day Mass Function (PDMF) with radial velocity anisotropies and then ultra-steep densities. In Section \ref{sec:other}, we discuss enhancements obtained with other proposed scenarios  and summarize our results in Section \ref{sec:summary}.

%We introduce the different kinds of close encounters that are relevant for our problem in \S \ref{sec:close}, and derive their analytical rates in \S \ref{sec:strong}.  In \S \ref{sec:loss}, we solve the Fokker-Planck equation in angular momentum space in the presence of close encounters. In \ref{sec:analytical} we present our analytical time-dependent solutions of the Fokker-Planck equation with strong scattering. In \S \ref{sec:results} we compute the impact of close encounters on TDE rates and discuss our findings for the E+A preference. In \S \ref{sec:conclusions}, we discuss the implications of our work and briefly summarize it. 

% $\rho \propto r^{-\gamma} $

%In \cite{}, we proposed a modified loss cone theory that takes into account not only weak scattering but also strong scattering. 
%Here we explore the different possible explanations for the green preference of TDEs taking into account both weak and strong scattering. 

\section{Loss cone theory}
\label{sec:LC}
\subsection{Classical loss cone theory}
\label{sec:2.1}
In NSC, stars and compact objects evolve over time due to two-body relaxation. In a spherical galaxy, the distribution function of stars $f(\bold{x},\bold{v})$ can be transformed by Jeans' theorem to $ f(\epsilon,J)$, where $\epsilon$ and  $J$ are the specific energy and angular momentum of a stellar orbit. Although stars diffuse in both energy $\epsilon$ and angular momentum $J$, for the near-radial orbits relevant for TDEs, the dominant and more rapid mechanism is the diffusion in angular momentum and we can write $f(\epsilon,J) = f_\epsilon(\epsilon) f_j(J)$.
%and assume a frozen distribution of energy. %we follow the one-dimensional Fokker Planck equation in angular momentum space. 
Assuming a frozen distribution of energy, stars are fixed in bins of orbital energy but are allowed to diffuse through angular momentum space through a random walk evolution. This process can be captured by the orbit-averaged Fokker-Planck equation (e.g. \cite{BahcallWolf76, MagorrianTremaine99}):
\begin{equation}
%\label{FK}
\frac{\partial f }{\partial \tau} = \frac{1}{4j}\frac{\partial}{\partial j } \left( j\frac{\partial f}{\partial j}\right) \label{eq:FP1}
\end{equation}
where $j \equiv J / J_{c}(\epsilon)= \mathcal{R}^{1/2}$ is a dimensionless angular momentum variable (normalized by the angular momentum of a circular orbit, $J_{\rm c}$), $\tau \equiv \mu(\epsilon) t \approx t/t_{\rm r}$ is a dimensionless time variable (normalized by the energy relaxation time $t_r$) and $\mu(\epsilon)$ the orbit-averaged diffusion coefficient at specific energy $\epsilon$:
\begin{equation}
\mu(\epsilon)=\frac{1}{P(\epsilon)} \oint \frac{d r}{v_{\rm r}} \lim _{\mathcal{R} \rightarrow 0} \frac{\left\langle(\Delta \mathcal{R})^{2}\right\rangle}{2 \mathcal{R}}. \label{eq:diffAvg}
\end{equation}
Here $P(\epsilon)$ is the orbital period of a radial orbit of energy $\epsilon$, $v_{\rm r}$ is the star's radial velocity, and the local diffusion coefficient $\langle (\Delta \mathcal{R} )^2\rangle$ is presented in Appendix \ref{app:diff}. 

%stars stellar profile $\gamma_\star=3/2$, $m_\star=1 M_\odot$, %monochromatic distribution, except when stated otehrwise, number of stellar-mass black hole,
%$\gamma_{\rm bh}=7/4$, $m_{\rm bh} = 15 M_\odot$
%Eddington integral

%For a stellar distribution with a density profile $\rho(r) = \rho_{\rm infl} (r/r_{\rm infl})^{-\gamma_\star}$, a Keplerian potential $\psi = GM_{\bullet} /r$ \footnote{Throughout the remainder of this paper, we adopt the usual stellar dynamics convention of positive-definite potentials and positive-definite energies for bound orbits.}

The stellar distribution function  $f_\star(\epsilon)$  is calculated using Eddington’s formula which can be simplified to:
\begin{equation}
    f_\star(\epsilon) = 8^{-1/2}\pi^{-3/2} \frac{\Gamma(\gamma_\star+1)}{\Gamma(\gamma_\star-1/2)} \frac{\rho_{\rm infl}}{\langle m_\star \rangle} \left( \frac{GM_{\bullet}}{r_{\rm infl}}  \right)^{-\gamma_\star} \epsilon^{\gamma_\star-3/2} \label{eq:DF}
\end{equation}
for an isotropic stellar distribution with a density profile $\rho(r) = \rho_{\rm infl} (r/r_{\rm infl})^{-\gamma_\star}$ and a Keplerian potential $\psi = GM_{\bullet} /r$. The (positive-definite) specific orbital energy is, for a given star at radius $r$ and velocity $v$, $\epsilon = \psi(r) - v^2/2$; $\langle m_\star \rangle$ is the average mass in the stellar population; and the radius of influence $r_{\rm infl}$ is defined as the radius that encloses a total mass of stars equal to the SMBH mass. 
\subsection{Impact of strong scattering}
\label{sec:2.2}
In addition to their numerous weak encounters, stars also have a much smaller number of strong encounters that are not taken into account in the classical Fokker-Planck equation. Those strong encounters are much less numerous than weak encounters but, as we have shown in \citeauthor{LC2}, they are efficient at removing stars on highly eccentric orbits, i.e., the stars that could have become TDEs. 
Let us briefly summarize the conditions for a star to be ejected. Let us consider a test star whose velocity is $\bold{V}$ while its velocity after a strong encounter becomes $\bold{V +  \delta v}$. Let $\theta $  be the angle between $\bold{V}$ and $\bold{\delta v}$, with $v_{\rm esc}$ the escape velocity at this point.  The star will be ejected if \citep{Henon}: 
%$$\lvert \bold{V + e} \rvert \geq v_{\rm esc} $$
\begin{equation}
\label{eq:conditionV1}
V^2 +  \delta v^2 + 2 V \delta v \cos \theta \geq v_{\rm esc}^2.
\end{equation}
This condition for ejection (Eq. \ref{eq:conditionV1}) remains the same for both strong encounters with equal mass scatterers and unequal mass scatterers. 

Assuming Keplerian motion and an escape velocity $v_{\rm esc}=\sqrt{2GM_\bullet/r}$, for equal mass scatterers, the local ejection rate writes, (\citeauthor{LC2}): 
\begin{equation}
\label{eq:ejectionrate}
 \dot{N}_{\rm ej}= \frac{ 2^{2 - \gamma_\star}  \pi \rho_{\rm infl} a^2 m_\star   V^{1 + 2\gamma_\star}  }{( 1+\gamma_\star) M^2} \left(\frac{GM_\bullet }{r_{\rm infl}} \right)^{ - \gamma_\star}
\end{equation}
with $V$ the local Keplerian velocity of the test star and $a$ its semimajor axis. 

Whereas for unequal mass scatterers, the local ejection rate becomes, (\citeauthor{LC2}): 
\begin{equation}
\begin{split}
\dot{N}_{\rm ej,u} = 16 \pi^2 G^2 m_{\rm bh}^2 \left( \int_{v_1}^{v_2} I_A f_{\rm bh}(v) v {\rm d}v  + \int_{v_2}^{v_3} I_B  f_{\rm bh}(v) v {\rm d}v +\right. \\ \left. \int_{v_3}^{v_{\rm esc}}  I_C f_{\rm bh}(v) v {\rm d}v \right) \label{eq:unequalEjectionsLocal}
\end{split}
\end{equation}
where 
\begin{equation}
\begin{split}
I_{A} &=\frac{2\left[v^{ 2}-\alpha\left(v_{\rm esc}^{2}-V^{2}\right)\right]^{3 / 2}}{3 V\left(v_{\rm esc}^{2}-V^{2}\right)^{2}} \\
I_{B} &=\frac{2\left[v^{ 2}-\alpha\left(v_{\rm esc}^{2}-V^{2}\right)\right]^{3 / 2}+v \left[2 v^{ 2}-3 \alpha\left(v_{\rm esc}^{2}-V^{2}\right)\right]}{6 V\left(v_{\rm esc}^{2}-V^{2}\right)^{2}}\\ 
&-\frac{\left(2 v_{\rm esc}+V\right)}  {6 V\left(v_{\rm esc}+V\right)^{2}}+ \frac{(1+\alpha)^{2}}{8 V\left(V+v\right) } \\
I_{C} &=\frac{3 v_{\rm esc}^{2}-V^{2}}{3\left(v_{\rm esc}^{2}-V^{2}\right)^{2}}+\frac{(1+\alpha)^{2}}{4\left(V^{2}-v^{2 }\right)}
\end{split}
\end{equation}
and the integration limits are given by:
\begin{equation}
\begin{split}
v_1 &= \sqrt{ \alpha (v_{\rm esc}^2 - V^2)} \\
v_2 &= \frac{1}{2} [ (1+ \alpha) v_{\rm esc} - (1- \alpha) V ]\\
v_3 &= \frac{1}{2} [ (1+ \alpha) v_{\rm esc} + (1 + \alpha) V ]
\end{split}
\end{equation}

As the orbital period is short compared to the relaxation time, local ejection rates per star can then be orbit averaged. The closed forms that we derived in \citeauthor{LC2} for some physically motivated values of $\gamma_{\rm bh}$ can be found in Appendix.\ref{app:analytics}. 
%The analytical solutions that we derived can be found in Sec. 3.1 of \cite{LC}.

%The orbit-averaged ejection rate does not have a general closed form.  
%However, for relevant values of $\gamma_\star$ which are integers or half-integers, an analytical solution exists.  Analytic solutions for some physically motivated values of $\gamma_\star$ are presented in Appendix \ref{app:analytics}.

\subsection{Modified Fokker-Planck equation}
\label{sec:2.3}
As we have shown, the ejection of stars from the distribution due to strong scatterings can be modeled by adding a sink term to the Fokker-Planck equation, which becomes (\citeauthor{LC2}): 
\begin{equation}
\frac{\partial f }{\partial \tau} = \frac{1}{4j}\frac{\partial}{\partial j } \left( j\frac{\partial f}{\partial j}\right) -  \frac{\langle \dot{N}_{\rm ej} \rangle}{\mu (\epsilon)}  f. \label{eq:FP2}
\end{equation}
where $\langle \dot{N}_{\rm ej}\rangle $ is the orbit-averaged rate of ejection due to strong scattering, $j $ and $\tau$ are dimensionless angular momentum and times as defined in the previous section. 
%\equiv J / J_{c}(\epsilon)=\mathcal{R}^{1 / 2}$ is a dimensionless angular momentum variable (normalized by the angular momentum of a circular orbit, $J_{\rm c}$). $\tau \equiv \mu(\epsilon) t \approx t/t_{\rm r}$ is a dimensionless version of time $t$.

The initial and inner boundary conditions depend on a dimensionless diffusivity parameter $q(\epsilon)=\mu(\epsilon)P(\epsilon)/j^2_{\rm lc}(\epsilon)$, which determines whether the loss cone evolves in the ``empty'' ($q \ll1$; stars immediately destroyed once $j\le j_{\rm lc}$) or ``full'' ($q \gg 1$; stars may move in and out of the loss cone multiple times per orbit) limits.

%where $j_{LC} = \sqrt{2GM r_t}/J_c(\epsilon)$ and $r_t$ the tidal disruption radius.

For an empty loss cone an absorbing boundary condition at the loss cone can be assumed while for a full loss cone, the distribution function only goes to zero at a much smaller value of dimensionless angular momentum, $j_0= j_{\rm lc}(\epsilon) \exp(-\alpha/2)$, where

\begin{equation}
    \alpha(q) \approx \left(q^2 + q^4 \right)^{1/4}
\end{equation}
is an approximate flux variable that smoothly bridges the empty and full loss cone limits 
%which has been derived by returning to the local (non-orbit-averaged) Fokker–Planck equation and determining how f varies with radial phase assuming $f = 0$ at periapsis 
\citep{CohnKulsrud78, Merritt13}. The presence of a sink term does not impact $\alpha$ as long as $j_o \ll j_{lc}$ (\citeauthor{LC2}). 
Hence the boundary conditions are:
\begin{equation}
f\left(j \leq j_{\rm 0}, t\right)=0 ;\left.\quad \frac{\partial f}{\partial j}\right|_{j=1}=0.
\end{equation}

%In both cases we take as an initial condition an isotropic distribution, $f(j)=1$, for $j_0 \le j \le 1$, and set $f(j)=0$ elsewhere.

% The intial and inner boundary conditions depend on a dimensionless diffusivity parameter $q(\epsilon)=\mu(\epsilon)P(\epsilon)/j^2_{\rm lc}(\epsilon)$, which is defined in terms of the size of the loss cone in dimensionless angular momentum space ($j_{\rm lc} \approx \sqrt{ 2GM R_{\rm t}}$).  The value of $q$ determines whether the loss cone evolves in the ``empty'' ($q \ll1$; stars immediately destroyed once $j\le j_{\rm lc}$) or ``full'' ($q \gg 1$; stars may move in and out of the loss cone multiple times per orbit) limits. 

The flux of stars that scatter into the loss cone per unit of time and energy is given by:
\begin{equation}
\mathcal{F}(t ; \epsilon)= 2 \pi^2 \mu(\epsilon) P(\epsilon) J_{\rm c}^2(\epsilon) f_\epsilon(\epsilon) \left ( j\frac{\partial f_j(j,t) }{\partial j}\right)_{j= j_{\mathrm{lc}}}. \label{eq:flux}
\end{equation}

Then, time-dependent TDE rates are obtained  by integrating $\mathcal{F}(\epsilon)$ across many bins of energy $\epsilon$, such that: 
\begin{equation}
    \dot{N}_{\rm TDE} (t)  = \int \mathcal{F}(t; \epsilon) {\rm d}\epsilon. \label{eq:totalRate}
\end{equation}

We derived analytical solutions of the modified Fokker-Planck equation with strong scattering Eq. \ref{eq:FP2} using the method of Frobenius (\citeauthor{LC2}). The analytical solutions for different slopes of scatterers can be found in Appendix \ref{app:sol}.

%The radial distribution of stars and compact objects evolves over time due to two-body relaxation. In the continuum limit, and assuming spherical symmetry, the stellar population can be represented with a distribution function $ f(\epsilon,J)$, where $J$ is the specific angular momentum of a stellar orbit. For the near-radial orbits relevant for TDEs, angular momentum relaxation is much faster than energy relaxation and we will use a two-timescale argument: we separate $f(\epsilon,J) = f_\epsilon(\epsilon) f_j(J)$ and assume a frozen distribution of energy. %we follow the one-dimensional Fokker Planck equation in angular momentum space. 

%We begin the integral at $\epsilon=\epsilon_{\rm infl}$ (as this is where our approximation of a Kepler potential breaks down, and in any case is usually around the peak of the flux curve), and integrate to a large value $\sim \epsilon_{\rm infl}$ (the exact upper limit has little impact on the results as the flux declines precipitously at high energies).  
%\section{Stellar properties proposed to explain the green preference}
\section{Radial velocity anisotropies}
\label{sec:anisotropies}
%The anisotropy parameter $\beta_a$ characterizes the extent to which  stellar orbits are predominantly radial or tangential.  
Velocity anisotropies i.e., the extent to which stellar orbits are predominantly radial or tangential have an influence on the number of stars getting tidally disrupted. Indeed, only stars with high eccentricities (low normalized angular momentum $j \leq j_{lc}$) can become TDEs. Hence, anisotropic distributions with more (respectively less) eccentric stars than the isotropic distribution would give rise to a higher (respectively lower) number of disrupted stars. 
%Observations have shown that, at least in our galaxy, young stars harbour a preference for tangential velocities \citep{}.  
\cite{Lezhnin} explored the impact of an anisotropic distribution with more tangential velocities and found that it could reduce the number of TDEs for a fraction of the relaxation time. \cite {Stone+18} proposed that the infall and tidal disruption of young massive clusters could give rise to an opposite distribution harboring a preference towards the radial component and found that, depending on the bias, TDE rates could be enhanced by a factor up to a few hundreds for up to the relaxation time.

Here, we explore the impact of such radial velocity anisotropies on TDE rate in the framework of our revised loss cone theory that takes into account both weak and strong scattering (\citeauthor{LC2}).
The anisotropy parameter $\beta_a$ characterizing the extent to which  stellar orbits are predominantly radial or tangential can be parametrized as: 
\begin{equation}
\beta_a \equiv 1 - \frac{T_{\perp}}{ 2 T_{\parallel} }
\end{equation}
where $T_{\perp}$ and $T_{\parallel}$ are the kinetic energies of tangential and radial motion respectively, with $\beta_a=1$ corresponding to all orbits being radial and $ \beta_a=0$ to an isotropic distribution. Hence, the initial condition for anisotropic velocity distributions writes:
\footnote{The condition is identical to the condition considered in \cite{Stone+18}}
\begin{equation}
f_j(j, t=0) = \frac{1-\beta_a}{1-j_{\rm{lc}} ^{2-2 \beta_a} } j^{-2 \beta_a}, \: \: \: j_{\rm{lc}}< j \leq  1 
\end{equation}

%The initial anisotropy parameter could, in principle, evolve with time (argument pk ce parametre devrait reduire). Hence considering a constant $\beta_a$ parameter results in finding an upper limit to the impact of radially-biased profile.
%maximum ROI 

\begin{figure}
\centering
\includegraphics[width=0.48\textwidth]{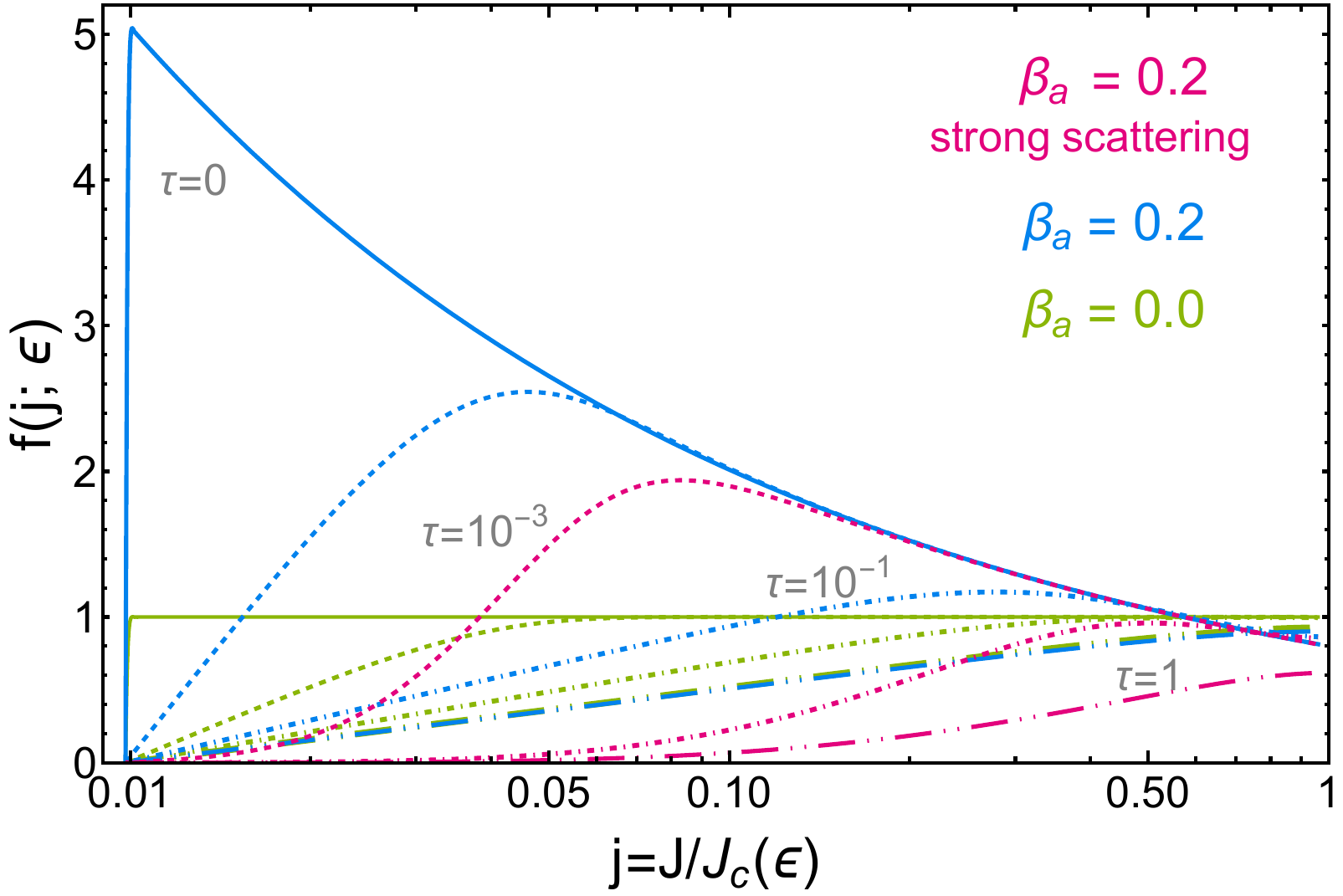}
\caption{Evolution of the distribution function $f(j;\epsilon)$ as a function of the dimensionless angular momentum j at fixed energy $\epsilon$, shown for different snapshots in dimensionless time $\tau = 0$ (solid), $\tau = 10^{-3}$  (dashed), $\tau = 10^{-1}$  (dot–dashed) and  $\tau = 1$  (dot–dot–dashed). Green lines correspond to an isotropic stellar cluster ($\beta_a = 0$). Blue lines show a case with moderate initial radial anisotropy ($\beta_a = 0.2$) without strong scatterings while pink lines show the same case with strong scatterings. All curves are shown in the empty loss cone regime with $q=0.006$ 
for a SMBH with $M_{\bullet} = 10^{6.5} M_{\odot}$
and stars with a power law density $\gamma_\star=3/2$. }
   \label{fig:radial}
\end{figure}

Fig.\ref{fig:radial} showcases the evolution of the distribution function at different times for a radially-biased distribution with $\beta_a \sim 0.2 $ with and without strong scattering as well as the evolution with an isotropic distribution. 

Without strong scattering, the change induced by the anisotropic distribution washes out after a relaxation time $t_r$ (i.e., $\tau=1$). However, when taking into account both weak and strong scattering, higher eccentricities stars (i.e., $j \leq 0.03$) are more depleted than the isotropic distribution at early times $ \tau \sim 10^{-3} \: t_r \sim 10^6 $ years. At later times $\tau \sim 0.1 \: t_r$, stars with all angular momentum are more depleted than in the case with an isotropic distribution.  As we have shown in \citep{LC}, strong scatterings are more efficient at depleting stars with the most radial orbits i.e. lower angular momentum. Hence, an anisotropic distribution with more stars with lower angular momentum induces a short-lived enhancement then followed by a stronger depletion of stars compared to the isotropic distribution. 

\begin{figure}
\centering
\includegraphics[width=0.47\textwidth]{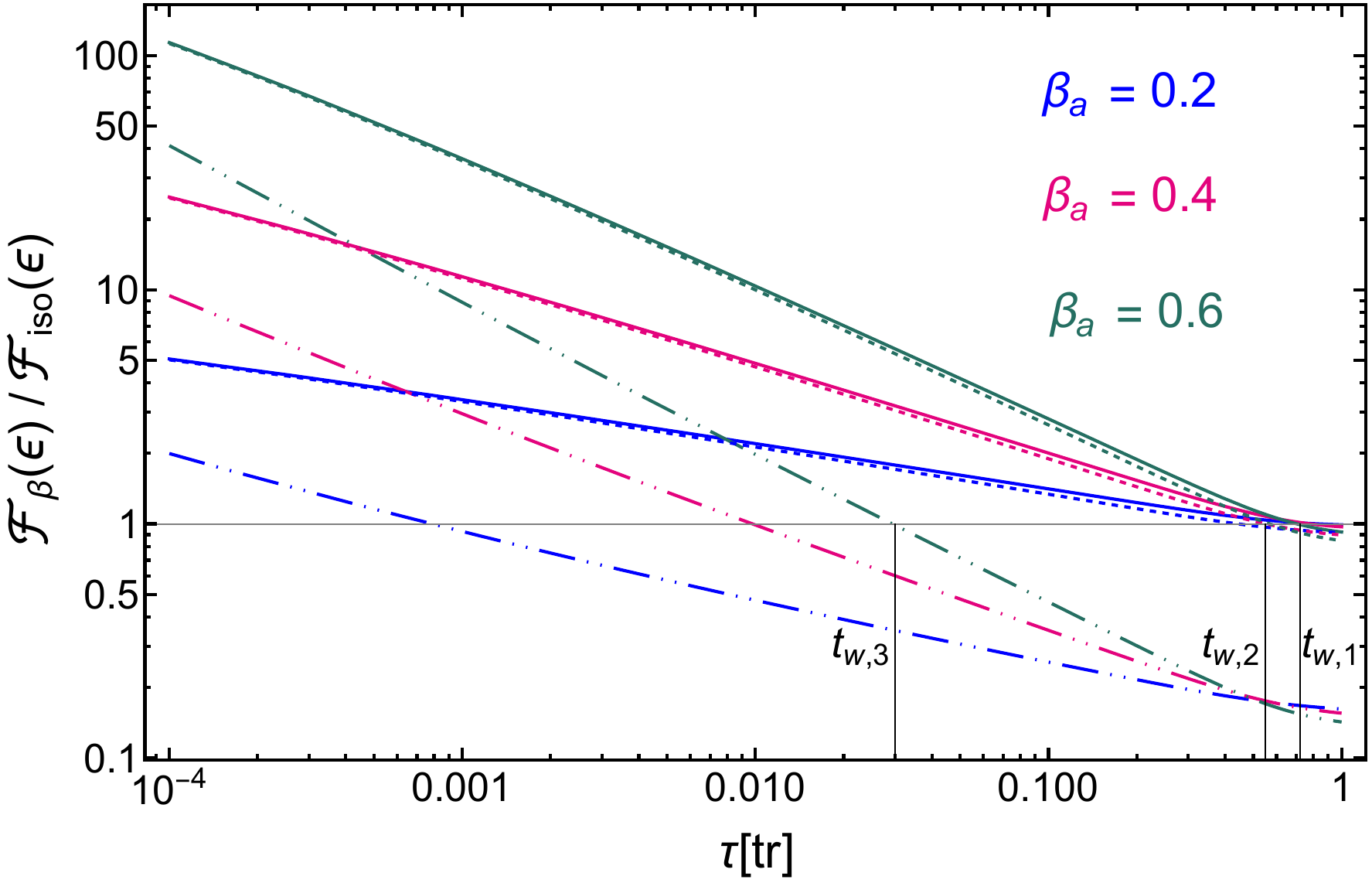}
 \caption{Time evolution of the flux of stars into the SMBH loss cone($\mathcal{F_{\beta}}$) normalized to the equivalent stellar flux for isotropic initial conditions, ($\mathcal{F}_{\rm iso}$) for a SMBH with $M_{\bullet} = 10^{6.5} M_{\odot}$ and a density power law of stars $\gamma_{\star} =3/2$. The horizontal axis is normalized to the relaxation time $t_r$. We consider three different anisotropy factors: $\beta_a = 0.2$ (blue), $\beta_a = 0.4$  (pink) and $\beta_a = 0.6$ (green). The full lines show the flux enhancements without strong scattering, the dashed lines depict the case of strong scattering $\gamma_{\rm bh}=7/4$, and the dot-dashed lines show the case of strong scattering and $\gamma_{\rm bh}=5/2$. The maximum times for the enhancements to completely wash out are also indicated: $t_{w,1}$ (without strong scatterings), $t_{w,2}$ (strong scatterings:  $\gamma_{\rm bh}=7/4$), $t_{w,3}$ (strong scatterings: $\gamma_{\rm bh}=5/2$). }
   \label{fig:flux}
\end{figure}

This effect is further explored in Fig.\ref{fig:flux} which presents the flux of stars into the loss cone for an anisotropic distribution normalized by the isotropic flux of stars where fluxes are obtained with Eq.\ref{eq:flux}. As expected, higher anisotropic factors give rise to higher enhancements. Without strong scatterings, depending on the anisotropy parameter $\beta_a$, fluxes are enhanced by a factor $5-100$. For all anisotropy parameters $\beta_a$, those enhancements wash out at $t_{w,1} \approx 0.7 \: t_r$. However, with strong scattering, enhancements and duration of those enhancements $t_w$ strongly depend on the stellar mass black hole density slope $\gamma_{bh}$. 
%In the case, and hence on strong segregation \citep{AlexanderHopman09}. 
In the case of weak segregation: $\gamma_{bh} =7/4$, the evolution of the enhancements induced by the different anisotropy factors is very similar to the evolution without strong scattering albeit the time for the enhancements to wash out is slightly smaller $t_{w,2} \approx 0.5 \: t_r$. For strong segregation and $\gamma_{bh} =5/2$, the fluxes are only enhanced by a factor $2-40$ at most and wash out early on at $t_{w,3} \approx 10^{-3}- 10 ^{-2} \: t_r$, depending on the anisotropy factor. It is important to note, that {\it enhancements turn into reduction for most of the relaxation time when accounting for strong scatterings. } 

%As shown, in Fig. \ref{fig:radial}, strong scatterings are more efficient at depleting the high-eccentricties stars that are overrepresented in anisotropic distributions. 

\begin{figure}
\centering
\includegraphics[width=0.49\textwidth]{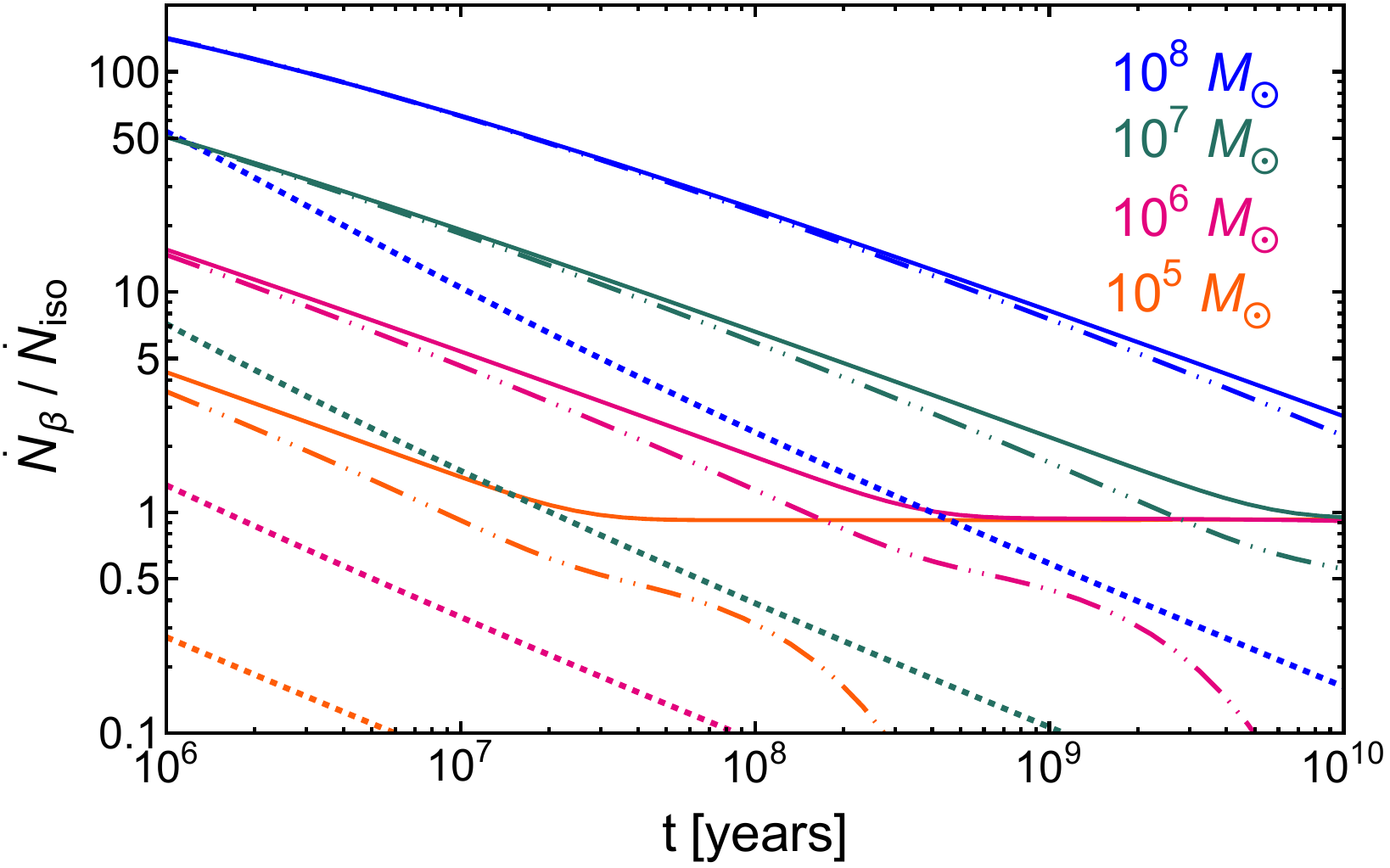}
 \caption{Evolution of TDE rate enhancements: $\dot{N}_{\beta}/\dot{N}_{\rm iso} $ for an anisotropy factor $\beta_a = 0.5$ as a function of post-starburst time $t$. Different colors correspond to different SMBHs as labeled in the figure. The full lines show the rate enhancements without strong scattering, the dashed lines depict the case of strong scattering and $\gamma_{\rm bh}=7/4$, and the dot-dashed lines show the case of strong scattering and $\gamma_{\rm bh}=5/2$.}
   \label{fig:rate}
\end{figure}

The fluxes of stars inside the loss cone are then integrated over a large number of energy bins following Eq.\ref{eq:totalRate}. Fig.\ref{fig:rate} showcases the evolution of TDE rate enhancements: $\dot{N}_{\beta} / \dot{N}_{\rm iso}$ for an anisotropy factor $\beta_a=0.5$,
chosen to be constant across all orbital energies $\epsilon$ and close to the maximum anisotropy factor, guaranteeing stability (e.g. \cite{Merritt85}). 

%rates of TDEs with an anisotropic distribution ($\dot{N}_{\beta_a}$ ) divided by the rates of TDEs for an isotropic distribution ($\dot{N}_{\rm iso} $) for an anisotropy factor $\beta_a=0.5$,

Without strong scattering, the enhancements induced by the anisotropic distribution depend on the SMBH mass: ranging from $4$ for a $M_\bullet =10^5 M_\odot$ to $150$ for a $M_\bullet=10^8 M_\odot$. The enhancements diminish with time such as $\dot{N}_{\beta_a} / \dot{N}_{\rm iso} \: \propto \: t^{-\beta_a}$ until reaching a plateau at $t_w$,
the time for enhancements to wash out. As shown in Fig. \ref{fig:flux}, $t_w$ is a fraction of the relaxation time and hence increases with the mass of the SMBH. 

With strong scattering, the evolution depends on the power-law density slopes of stellar mass black holes $\gamma_{\rm bh}$. For $\gamma_{\rm bh}=7/4$, the evolution is similar to the one without strong scattering for higher SMBH masses. For smaller SMBH masses, the time for enhancements to wash out $t_w$ is small enough that enhancements turn into reductions after $ \sim 10^7$ Myr for $M_\bullet = 10^5 M_\odot$ and  $ \sim 10^8$ Myr for $M_\bullet = 10^6 M_\odot$. However for $\gamma_{\rm bh}=5/2$, strong scatterings are so efficient that, even after 1 Myr, the enhancements induced by the anisotropic distribution have already washed out for $M_\bullet \leq 10^6 M_\odot$ and are a factor $50$ for a $M_\bullet = 10^8 M_\odot$. It can also be noted that enhancements diminish at a slightly higher rate and hence after 20 Myr, {\it enhancements have turned into reductions} for SMBH up to $M_\bullet = 10^7 M_\odot$ and are a mere factor $6$ for $M_\bullet = 10^8 M_\odot$. 
%and the impacts of the anisotropy factor $\beta_a$ on TDE rates are showcased in In Fig. \ref{fig:rate}. 

\section{Ultra-steep stellar densities}
\label{sec:overdensities}

%In a relaxed NSC, stars are expected to settle to the well-known Bahcall–Wolf cusp $\gamma_\star =7/4$ or to weaker cusp $\gamma_star \approx 1.3 -1.5 $  for a sytem composed of both stars and compact objects. 

%A relaxed NSC star population would have a density slope $\gamma_\star =7/4$ in the absence of heavier objects while it would assume $\gamma_\star \approx 1.3 -1.5 $ in their presence (e.g., \cite{BahcallWolf76, BW77}). 
In a relaxed nuclear star cluster (NSC), stars are expected to settle to the well-known Bahcall–Wolf cusp $\gamma_\star =7/4$ in the absence of heavier objects while a weaker cusp $\gamma_\star \approx 1.3 -1.5 $ is expected in their presence (e.g., \cite{BahcallWolf76, BW77}).
Ultra-steep profile with $\gamma_\star \approx 2.25- 2.5$ could be formed in the case of a very centrally concentrated (i.e. inside the sphere of influence) star formation \citep{Generozov+18}. Moreover, \cite{Young} showed that the stellar slope may assume steeper values of $\gamma_\star \gtrsim 2$, for an adiabatically growing SMBH.

%However, changing the overall slope of the NSC requires the majority of the stars to form such a steep slope over a short time. Whether this is realistic is debatable. Nevertheless, here we assume initial conditions with such steep slopes exist and test the possibility that they can indeed enhance the TDE rates when accounting for strong scattering.

\cite{Stone+18} found that ultra-steep density profiles with $\gamma_\star \approx 2.25 - 2.5$ could enhance TDE rates by a factor $\approx 10-100$. As we showed that the efficiency of strong scattering highly depends on the density slopes of the scatterers, we explored the effect of ultra-steep density profiles of stars considering the effect of strong scattering at early times when the erosion of the cusp was negligible (\citeauthor{LC2}). We found that strong scattering could reduce the enhancements induced by ultra-steep densities (Fig. 10, \citeauthor{LC2}). Motivated by these findings, we further explore this effect here considering the effect of strong scattering from both stars and stellar mass black holes. We also account for the erosion of the cusp and hence can compute the evolution with time of ultra-steep stellar densities with strong scatterings. 
%at all times (i.e., including after the erosion of the cusp). 

Assuming a stellar density slope $\gamma_\star$, stellar-mass black holes stemming from such a distribution of stars will acquire a density slope $\gamma_{\rm bh}$ such that $\gamma_{\rm bh} \geq \gamma_\star$. Hereafter, to be conservative, we shall assume that $\gamma_{\rm bh} = \gamma_\star$. Ultra-steep stellar density profiles will result in much shorter relaxation times, especially at small energies. An ultra-steep cusp erodes from the inside out and the ultra-steep density evolves from a power law $ \rho \propto  r^{-\gamma_{\rm steep}}$ to a broken power law:

\begin{equation}
\label{Eq:density}
\rho(r,t)=\left\{\begin{array}{ll}
\rho_{\rm inf} (r/ r_{\rm inf}) ^{-\gamma_{\rm steep}},  & r > r_{\rm b} (t) \\
\rho_{\rm inf} (r/ r_{\rm b}(t)) ^{-\gamma_{\rm rel}} (r_{\rm b}(t)/ r_{\rm inf}) ^{-\gamma_{\rm steep}},  & r \leq r_{\rm b} (t) 
\end{array}\right.
\end{equation}
where $\gamma_{\rm steep}$ is the initial ultra-steep value of the cusp while $\gamma_{\rm rel}$ is the relaxed value, $r_{\rm b}, is$. We compute the evolution of the density slopes of the stellar mass black holes $\gamma_{\rm bh}(t)$ and stars $\gamma_\star(t)$ by solving the one-dimensional time-dependent Fokker-Plank equation in energy with the code PHASEFLOW developed by \cite{Vasiliev17}.  

\begin{figure}
\centering
\includegraphics[width=0.49\textwidth]{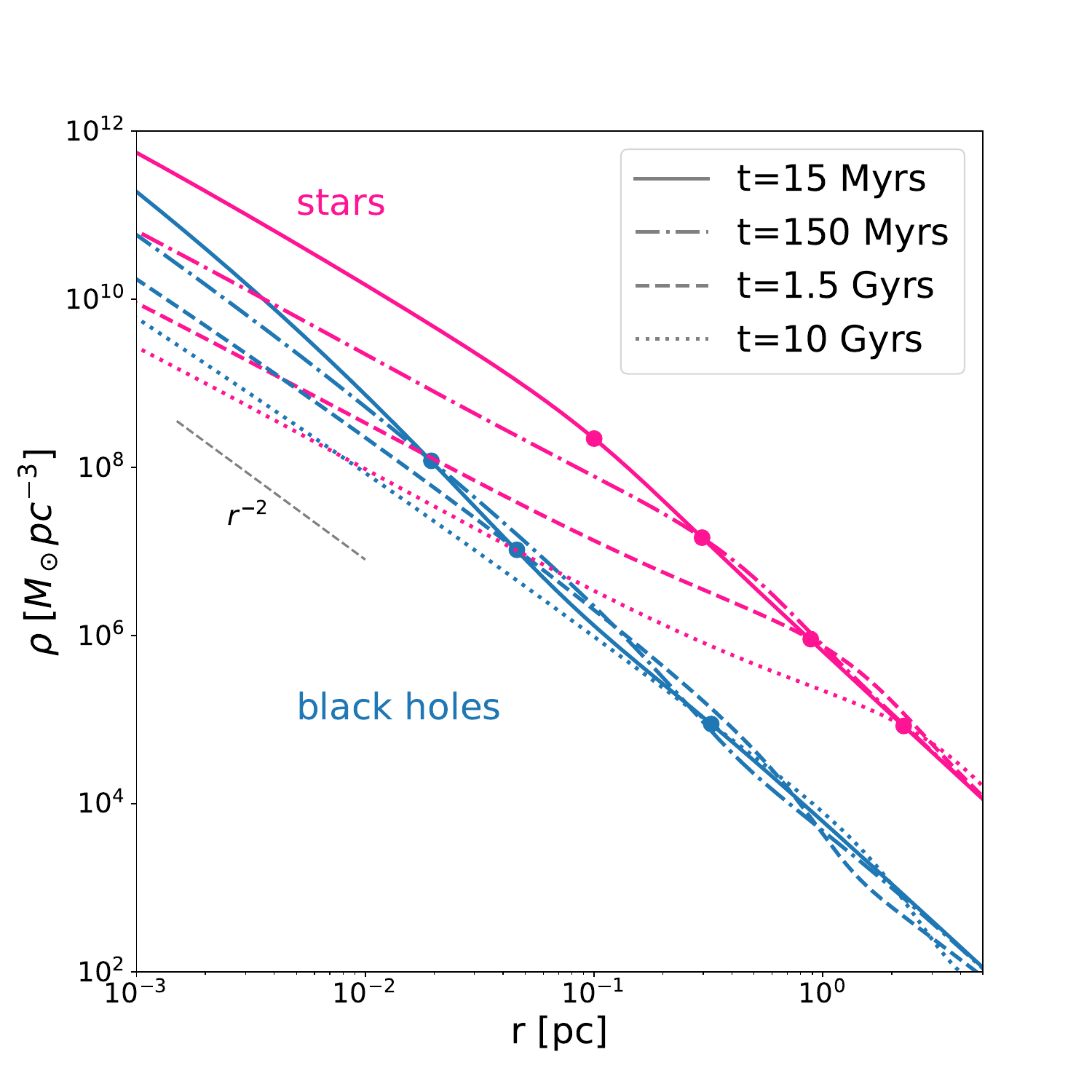}
 \caption{Evolution of an initially ultrasteep profile of stars and stellar mass black holes: $\rho \propto r^{-\gamma}$, $\gamma_\star=\gamma_{\rm bh} =5/2$ using the PHASEFLOW code. Here we consider a SMBH with mass $M= 10^{6.5} M_\odot $. Pink lines showcase the evolution of stars while blue lines correspond to stellar-mass black holes. The different lines correspond to post-starburst ages $t= 15$ Myrs (solid),  $t= 150$ Myrs (dot-dashed), $t= 1.5$ Gyrs (dotted) and $t= 10$ Gyrs (dotted). The dots mark the break radius $r_{\rm b}$ at different times for both star and stellar mass black holes.}
   \label{fig:gamma}
\end{figure}

Fig. \ref{fig:gamma} shows the evolution of the densities slopes $\gamma_\star(t)$ and $\gamma_{\rm bh}(t)$ at different times for a SMBH $M_\bullet = 10^{6.5} M_\odot$ for an initial ultra-steep profile $\gamma_\star=\gamma_{\rm bh} =5/2$. After the erosion, the power law density becomes a broken power law as defined in Eq. \ref{Eq:density}. The break radius for both stars and stellar-mass black holes are indicated by the dots for different times. The stellar-mass black holes erode more slowly than the stars: at $t= 15 $ Myr, $r_{\rm b, \star} \sim 10^{-1} $ parsecs while $r_{\rm b, bh} < 10^{-3} $ parsecs. 

%The evolution of the break radius for stars is 

%After t, the star density becomes a power law as defined in Eq. \ref{Eq:density} with a small break radius $r_b \approx  $. However after t, $r_b \approx  r_{\rm inf}$. On the other end, the slope of the stellar mass black holes remains approximately constant such as $\gamma_{\rm bh}(t) \approx \gamma_{\rm bh, 0}$. 
%The evolution of the break radius $r_b (t)$ has crucial implications when taking into account strong scattering. Indeed, as we have shown in \cite{LC}, strong scatterings are pericenter dominated, and negligible for shallow density profile of the stars $\gamma_\star \approx 3/2$.

Considering a broken power-law density has implications at different steps of the calculation. Firstly, for such a broken power-law, making use of the Eddington's formula the density function becomes: 
\begin{equation}
\begin{split}
    f(\epsilon, t) = 8^{-1/2}\pi^{-3/2} \frac{\Gamma(\gamma_{\rm rel}+1)}{\Gamma(\gamma_{\rm rel}-1/2)} \frac{\rho_{\rm infl}}{\langle m_\star \rangle}  \left( \frac{GM_{\bullet}}{r_{\rm b}(t)}   \right)^{-\gamma_{\rm rel}}  
   \\ \times   \left( \frac{r_{\rm b}(t)}{r_{\rm infl}}  \right)^{-\gamma_{\rm steep} }  \epsilon^{\gamma_{\rm rel}-3/2} 
    \label{eq:DF_broken}
    \end{split}
\end{equation}

%As $\mu(\epsilon)$ depends on the density function (see Eq. ), the becomes time-dependent: $\mu(\epsilon, t) \propto r_{b,\star} (t) ^ {-\gamma_{\rm steep, \star} + \gamma_{\rm rel, \star}}$
This change on the density function impacts the orbit-averaged diffusion coefficient $\mu(\epsilon)$ which becomes time-dependent:  $\mu(\epsilon, t) \propto r_{\rm b,\star} (t) ^ {-\gamma_{\rm steep, \star} + \gamma_{\rm rel, \star}}$. 

Local ejection rates for strong scattering are also modified when taking into account a broken power-law and become time-dependent. For equal mass scatterers, we find that the local ejection rate becomes: 
\begin{equation}
\begin{split}  
 \dot{N}_{\rm ej} (t)= \frac{ 2^{2 - \gamma_{\rm rel}}  \pi \rho_{\rm infl} a^2 m_\star   V^{1 + 2\gamma_{\rm rel}  }}{( 1+\gamma_{\rm rel}) M^2} 
\\ \times \left(\frac{GM_\bullet }{r_{\rm b} (t)} \right)^{ - \gamma_{\rm rel}} \left(\frac{r_{\rm b} (t) }{r_{\rm infl}} \right)^{ - \gamma_{\rm steep}}
 \end{split}
 \label{eq:ejectionrate_broken}
\end{equation}

Interestingly, the average ejection rate for equal mass scatterer has the same time dependence as the orbit-averaged diffusion coefficient $\mu$. Hence, the sink term for equal mass scatterers does not depend on time. However, for unequal mass scatterers, the unequal ejection rate has the following time-dependence $ \dot{N}_{\rm ej} (t) \propto r_{\rm b, bh} (t)^ {-\gamma_{\rm steep, bh} + \gamma_{\rm rel, bh}} $, while the orbit-averaged diffusion time dependence remains the same $\mu(\epsilon, t) \propto r_{\rm b,\star} (t) ^ {-\gamma_{\rm steep, \star} + \gamma_{\rm rel, \star}}$. Hence, the unequal sink term becomes time-dependent for broken-power law distributions. 

%the orbit-averaged diffusion $\mu$ remains the same  the ejection rate for unequal mass scatterer is such also becomes time-dependent (), where rb this time depending on the erosion of stellar-mass black holes. Hence the sink term is time-dependant.  

%Hence, for a given semi-major axis $a$ (respectively energy $\epsilon$) if the pericenter is smaller than the break radius: $r_p \leq r_b(t)$, the impact of strong scattering from other stars will be negligible. Here we compute the minimum semi-major axis $a_m$ for which both strong scattering of stars and stellar mass black holes should be taken into account:
%\begin{equation}
  %a_m (t) = \frac{r_b(t)}{ 1- \sqrt{(1-j_{o}^2)}} 
%\end{equation}

To compute TDE rates for ultra-steep profiles, we extract the break radius $r_{\rm b}(t)$ for stars and stellar mass black holes from PhaseFlow \citep{Vasiliev17}. Then, we compute both the orbit-average diffusion coefficient $\mu (\epsilon, t)$ and the sink term accounting for ejections from strong scatterings ( Eq.\ref{eq:DF_broken}- \ref{eq:ejectionrate_broken}). Finally, we integrate the Fokker Planck equation Eq.\ref{eq:FP2} with our time-dependent sink terms. 
%using a sink term that take into account strong scattering from both stars and stellar mass black holes for times and semi-major axis that are characterized by $a > a_m(t)$, while we use the sink term accounting only from scattering from stellar mass black holes other times and semi-major axis. 
%Fig.\ref{fig:flux_2} showcases the ratio of the flux of stars with an ultra-steep distribution divided by the flux of stars for a shallow slope $\gamma_\star =3/2$ where fluxes are obtained with Eq.\ref{eq:flux}.

\begin{figure}
\centering
\includegraphics[width=0.49\textwidth]{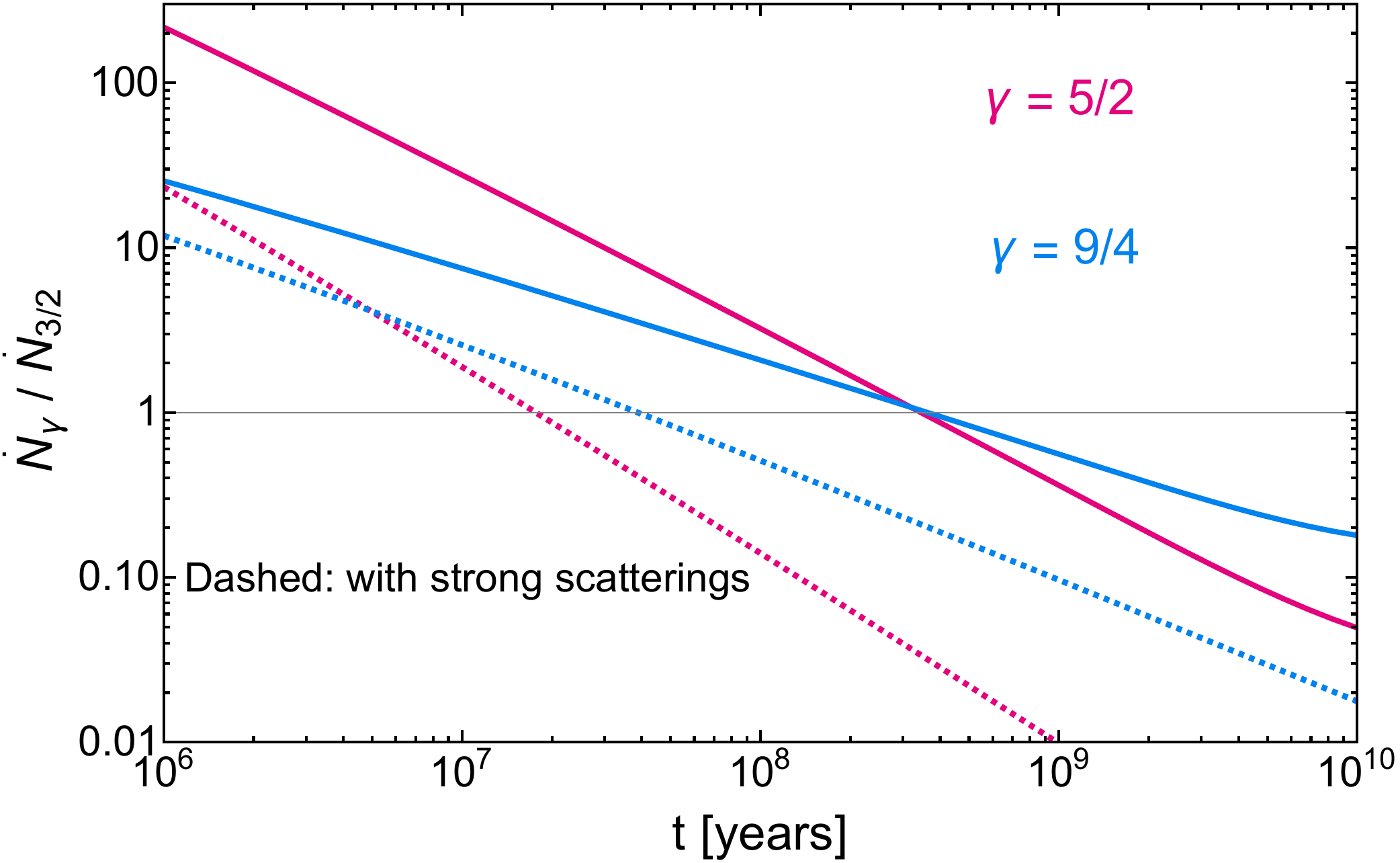}
 \caption{Evolution of TDE Rate enhancements: $\dot{N}_{\gamma}/\dot{N}_{3/2} $ for a SMBH $M_\bullet = 10^{6.5} M_\odot$ as a function of post-starburst time $t$. 
 Blue lines correspond to an initial ultra-steep profile of both stars and black holes of $\gamma=9/4$ while pink lines correspond to $\gamma=5/2$. The full lines show the rate enhancements without strong scattering while dashed lines show the evolution with strong scattering. The densities of both stars and stellar mass black holes have been evolved with the code Phaseflow, see Fig. \ref{fig:gamma}.}
   \label{fig:rate_ultrasteep}
\end{figure}

%The fluxes of stars inside the loss cone are then integrated over a large number of energy bins following Eq. \ref{eq:totalRate}. 
Fig.\ref{fig:rate_ultrasteep} showcases the evolution of TDE rate enhancements $\dot{N}_\gamma/\dot{N}_{3/2} $  for different ultra-steep profiles where rates are obtained by integrating over a large number of energy bins following Eq.\ref{eq:totalRate}, with and without strong scattering.
%with time of TDE rates with an ultra-steep stellar distribution ($\dot{N}_\gamma$) divided by the TDE rates for a relaxed stellar distribution ($\dot{N}_{3/2} $), where rates are obtained by integrating over a large number of energy bins following Eq.\ref{eq:totalRate}, with and without strong scattering. 
Without strong scatterings, TDE rates for ultra-steep densities of stars and stellar mass black holes are enhanced by up to a factor $\sim 200$ for $\gamma=5/2$ and $\sim 25$ for $\gamma=9/4$. %Those enhancements decrease with time such as $\dot{N}_\gamma \propto r_b (t)^{2\: (3/2 -\gamma)}$, the lines corresponding are shown Fig.\ref{fig:rate_ultrasteep}. 
Those enhancements decrease with time: such as the higher the density slope $\gamma$ the quicker the decrease. They wash out after a few hundreds Myrs for both profiles and are replaced by small reductions of TDE rates at later times. 
However when taking into account strong scatterings, 
%{\it Strong scatterings completely change the fate of ultra-steep profiles}, as 
enhancements are at most $\sim 20 $ %HBP - missing number
for $\gamma=5/2$ and $\sim 10 $ for $\gamma=9/4$. Enhancements only last a few tens Myrs and then {\it turn into significant reductions for most of the relaxation time}. 
%$ After a few tens of million years, enhancements induced by ultra-steep profiles are replaced by reductions. Moreover, at later times $t> 3 10^7 $ yr,  the ultra-steep profile induces a reduction in the TDE rates when accounting for strong scatterings. 

%For the curves accounting for strong scattering, depending on the criterion given by Eq. \ref{Eq}

\section{Combination of different present-day mass-functions with stellar properties}
\label{sec:IMF}
%Even though the monochromatic distribution has been extensively considered in literature, 

A present-day mass function $dN/dm_\star$ (PDMF) of stars is a more realistic representation of an NSC star population than the extensively considered monochromatic distribution. Hence, in sections Sec. \ref{sec:anisotropies} and Sec. \ref{sec:overdensities} we considered a Kroupa PDMF \citep{Kroupa}. 

Simplified PDMFs can be derived from IMFs by setting a cut-off mass to the IMF (Eq.\ref{eq:IMF}). Here, the cut-off mass is defined as $m_{max}= 2 M_\odot$ corresponding to a stellar population of age $\sim 1.8 $ Gyr.

Observations have shown that the young stellar population in our galactic center exhibit a top-heavy IMF (e.g., \citep{Bartko, Lu}. %HBP add citetion.
Moreover, \cite{Bortolas} suggested that top-heavy initial IMFs could slightly increase TDE rates. Motivated by this, we explore the effects of combining different PDMFs with either velocity anisotropies or ultra-steep stellar densities. 

All IMFs are defined by: 
\begin{equation}
\label{eq:IMF}
\chi (m) \propto \left\{\begin{array}{ll}
m_{\star}^{-1.3},  & m_{\star}<0.5 M_{\odot} \\
m_{\star}^{-\alpha}, & m_{\star} \geq 0.5 M_{\odot} 
\end{array}\right.
\end{equation}
with $\alpha=2.3$ for Kroupa and $\alpha=\{1.5, 1.7, 1.9 \}$
for top-heavy IMFs.
Following \cite{MagorrianTremaine99}, we also apply a truncation for the smallest masses such as $m_{\rm min}= 0.08 M_\odot$. 

In addition to stellar properties, the presence of compact objects also impacts the TDE rates, as their presence enhances the angular momentum diffusion coefficients $\mu$ (Eq.\ref{eq:diffAvg}). Throughout this paper, we consider that stars account for 97\% of the total mass while stellar mass black holes account for 3\% of the total mass and have a mass $m_{\rm bh}= 15 M_\odot$.  

%Stellar radius small impact 

\begin{figure}
\centering
\includegraphics[width=0.49\textwidth]{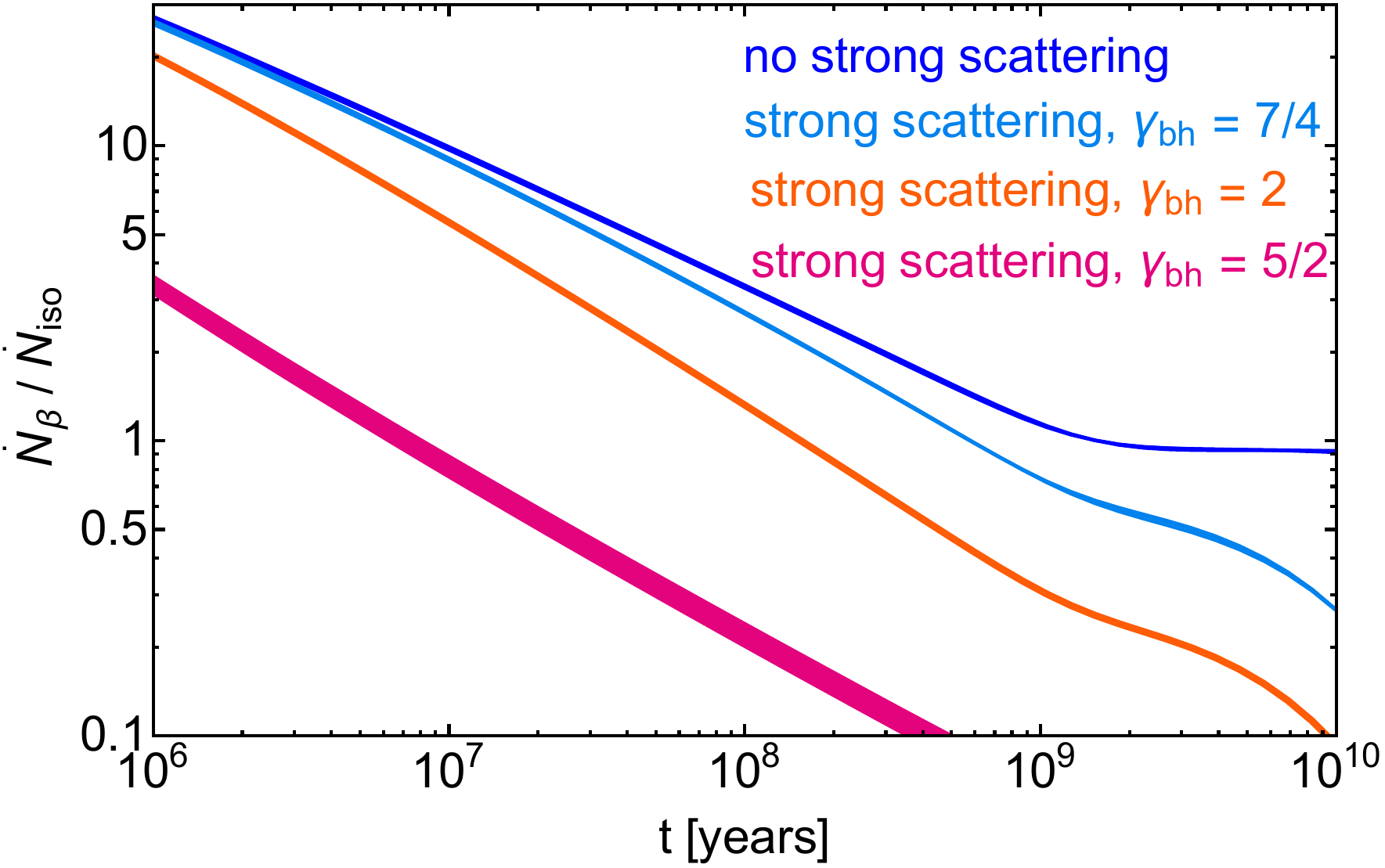}
 \caption{Evolution of TDE Rate enhancements $\dot{N}_{\beta}/\dot{N}_{\rm iso} $ for an anisotropy factor $\beta_a = 0.5$ as a function of post-starburst time $t$. As labeled the different colors correspond to the following different cases: no strong scatterings (blue), strong scatterings with $\gamma_{\rm bh}=7/4$ (light blue), strong scatterings with $\gamma_{\rm bh}=2$ (orange) and strong scatterings with $\gamma_{\rm bh}=5/2$ (pink).  The impact of different PDMFs is small and shown in the thickness of the different lines. For all curves, we consider a SMBH mass $M_\bullet = 10^{6.5} M_\odot$ and a density power law of stars $\gamma_\star =3/2$.
}
   \label{fig:rate_PDMF}
\end{figure}

%As can be seen in Fig.\ref{fig:rate_PDMF}, the evolution of TDE rates when combining different top-heavy PDMFs with velocity anisotropies is very similar to the evolution obtained for a Kroupa PDMF combined with velocity anisotropies Fig.\ref{fig:rate} . 
Fig.\ref{fig:rate_PDMF} showcases the evolution of TDE rates enhancements $\dot{N}_{\beta}/\dot{N}_{\rm iso} $ obtained when combining different PDMFs with an anisotropic distribution. The anisotropy factor $\beta_a=0.5$ is chosen to be the same as in Fig.\ref{fig:rate} (a value close to the maximum anisotropy factor guarantying stability (e.g. \cite{Merritt85}). The impact of combining different PDMFs with stellar anisotropies is shown in the thickness of the different lines as is minor for all cases. The effect slightly increase when considering strong scatterings and a stellar-mass black hole slope $\gamma_{\rm bh}=5/2$. After $10$ Myr, enhancements induced by a high anisotropy factor $\beta_a =0.5$ have completely washed out for a steep profile of stellar mass black holes $\gamma_{\rm bh}=5/2$ and are a mere factor $5-8$ for shallower profiles of stellar mass black holes $\gamma_{\rm bh}=7/4-2$. TDE rates enhancements when combining different PDMFs with an anisotropic distribution are very similar to enhancements obtained for an anisotropic distribution.

%ratio of TDE rates with an ultra-steep density ($\dot{N}_{\gamma}$) combined with different PDMFs divided by the TDE rates for a relaxed Kroupa distribution ($\dot{N}_{3/2} $).

\begin{figure}
\centering
\includegraphics[width=0.49\textwidth]{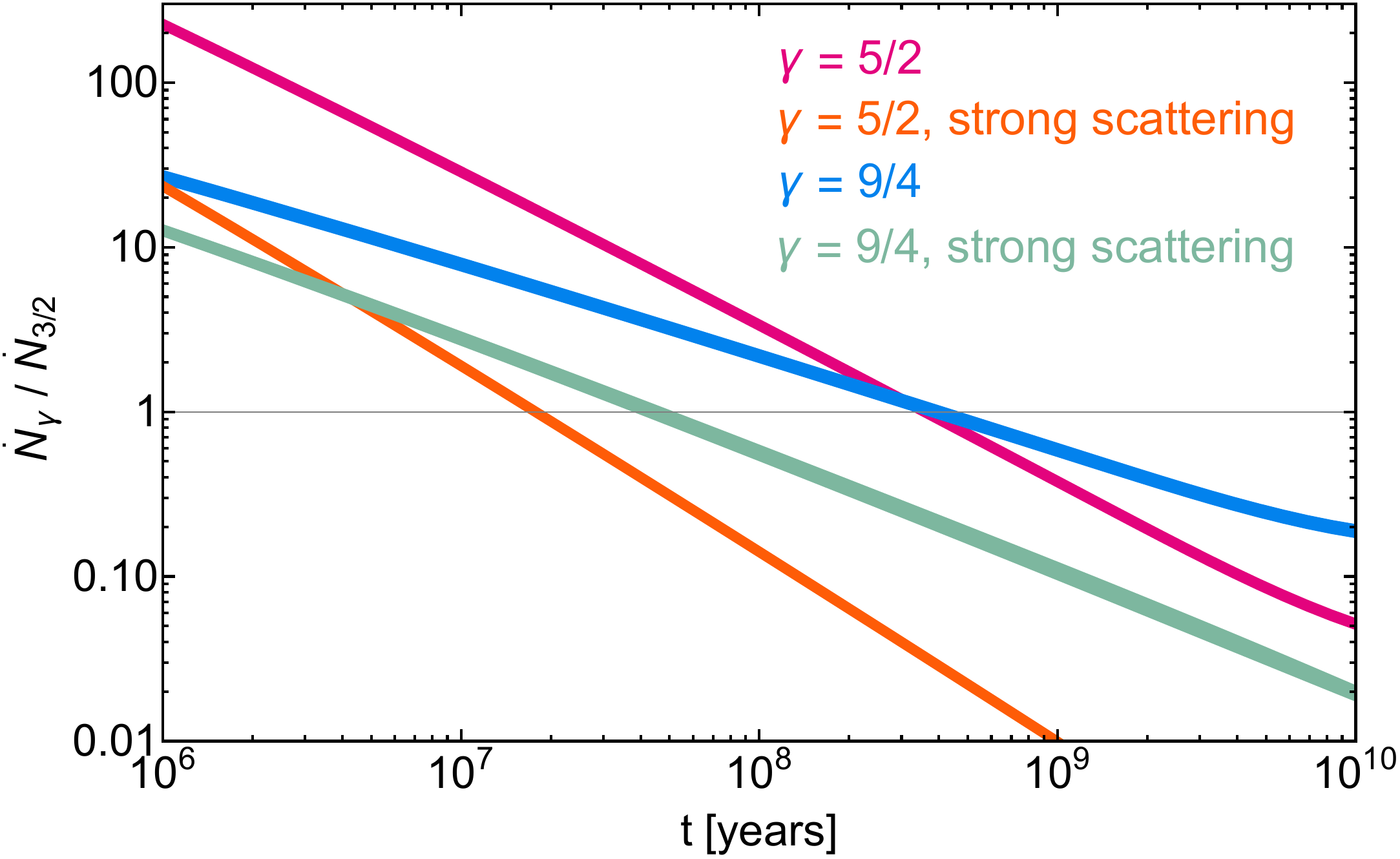}
 \caption{Evolution of TDE Rate enhancements  $\dot{N}_{\gamma}/\dot{N}_{3/2} $ for ultra-steep profiles combined with different PDMFs as a function of post-starburst time $t$. As labeled the different colors correspond to different initial ultra-steep profiles: $\gamma =5/2$ without strong scatterings (pink),  $\gamma =5/2$ with strong scatterings (orange), $\gamma = 9/4$ without strong scatterings (blue),  $\gamma = 9/4$ with strong scatterings (green). The densities of both stars and stellar mass black holes have been evolved with the code Phaseflow, see Fig. \ref{fig:gamma}.  The impact of different PDMFs is small and shown in the thickness of the different lines. For all curves, we consider a SMBH mass $M_\bullet = 10^{6.5} M_\odot$.}
   \label{fig:rate_PDMF_2}
\end{figure}

Fig.\ref{fig:rate_PDMF_2} showcases the evolution of TDE rates enhancements $\dot{N}_{\gamma}/\dot{N}_{3/2} $ when combining different PDMFs with ultra-steep density profiles. As in Fig.\ref{fig:rate_PDMF}, the impact of combining different PDMFs with stellar anisotropies is shown in the thickness of the different lines and is minor for all cases. We find that combining different PDMFs with ultra-steep density profiles results in a few percents changes in enhancements. Hence, we find that combining different PDMFs with either radial velocity anisotropies or ultra-steep stellar densities cannot explain the post-starburst preference of TDEs when accounting for strong scatterings.  
\\
\\

%, stemming from the fact that different PDMFs induce a slightly different orbit average coefficient enters $\mu$ (Eq.), which enters the TDE rates through both the flux of disrupted stars (Eq.) and the sink term (Eq.) in the presence of strong scatterings.   

%Without strong scattering, the differences between PDMFs are so small that the differences are undistinguishabel. 
 %In the presence of strong scatterings, the differences are still very small, maximum () but not undistinguable. This stems from the fact that a different PDMF induce a slighlty different orbit average coefficient enters $\mu$ (Eq.), which enters the TDE rates through both the flux disrupted stars (Eq.) and the sink term (Eq.) in the presence of strong scatterings.   

%\subsection{Importance of the age of the starburst}

%\section{Disk and MBH properties}

%\section{Observational implications}

\section{Alternative models for TDE enhancement rates}
\label{sec:other}
Given our results which challenge the leading suggested scenarios for TDE rate enhancements, it is important to briefly review other suggested scenarios.

\subsection{Massive Perturbers and Nuclear Spiral Arms}
Before the discovery of the post-starburst preference, \citet{Perets+07a, Perets:2008} proposed that massive perturbers, such as giant molecular clouds, could significantly reduce two-body relaxation times and enhance TDE rates. Nuclear spiral arms might have a similar effect \citep{Hamers:2017}. While these mechanisms could preferentially increase TDE rates in gas-rich and post-merger galaxies (which might have preferentially more molecular clouds serving as perturbers), by a small factor, they are unlikely to produce the order of magnitude or more enhancements observed in post-starburst galaxies. 

\subsection{Binary massive black holes}
The first proposed explanation for the post-starburst preference invoked the correlation between starbursts and galaxy mergers \citep{Arcavi+14}.  If many post-starburst galaxies are also post-merger galaxies, their nuclei may contain SMBH binaries which can increase TDE rates by many orders of magnitude (relative to galactic nuclei with solitary SMBHs) through a combination of Kozai cycles \citep{Ivanov+05} and chaotic three-body scatterings \citep{Chen+11, Wegg&Bode11}.  However, even though SMBH binaries may temporarily enhance TDE rates by multiple orders of magnitude, the short timescales for such enhancements (typically $\sim 10^5~{\rm yr}$, e.g. \citealt{Wegg&Bode11}) may make it challenging for this mechanism to explain the global fraction of all TDEs seen in post-starburst galaxies (see also discussions in \citealt{StoneMetzger16, Saxton+18}).

\subsection{AGN disk} 
The influence of an AGN disk on TDE rate has been first explored by \citep{Kennedy} who found enhancements of TDE rates by a factor $\sim 10$. Such a factor cannot explain alone the amplitude of observed enhancements. Recently, \cite{Wang24} explored the "wet" channel for TDEs and found that, enhancements could reach 2 order of magnitude for a very high star formation combined with a high viscosity and a high efficiency (i.e., conversion of the rest mass energy from star formation into radiation). However, such extreme environments require untypical conditions.   

%that could only be sustained for $\sim $ yr,  %HBP - is there a missing number?
%and are irrelevant on long timescales.  

\subsection{Secular Effects in Eccentric Stellar Disks} 
Nuclear starbursts can potentially generate eccentric stellar disks where secular effects dramatically increase TDE rates \citep{Madigan+18, Wernke}. However, this mechanism requires a relatively small nuclear cluster mass to avoid quenching coherent secular evolution through mass precession. 
This condition may be problematic for explaining the post-starburst preference, as most low-mass SMBHs coexist with substantial nuclear star clusters. The ideal environment for this mechanism —a disk-dominated nuclear stellar population—is more likely in massive galaxies with SMBHs ($M_\bullet \gtrsim 10^8$ M$_\odot$), which can only disrupt higher mass of post-main-sequence stars \cite{Antonini+15} and account for a small fraction of observed TDEs (e.g. \citep{Yao+23})

\section{Summary} 
\label{sec:summary}
We studied the scenarios invoking stellar properties to explain the post-starburst preference of TDEs in the framework of our revised loss cone theory that takes into account both weak and strong scatterings (\citeauthor{LC2}). We showed that enhancements induced by radial velocity anisotropies depend both on the anisotropy factor $\beta_a$ and on the mass of the SMBH: the more massive the SMBH the greater the increase and the longer its duration. When taking into account strong scatterings, we found that radial velocity anisotropies could not explain the post-starburst preference of TDEs except for the rare case of a high mass SMBH $M_\bullet \sim 10^8 M_\odot$ combined with a high radial anisotropy $\beta_a \sim 0.5$  and a shallower profile of stellar-mass black holes $\gamma_{\rm bh} =7/4$. 

We showed that ultra-steep stellar densities with $\gamma_\star \geq 9/4$ could enhance TDE rates by a factor $\sim 20-200$ depending on the density slope, without strong scatterings. However, when taking into account strong scatterings, enhancements induced by ultra-steep stellar densities i) are at most a factor $\sim 10-20$ at very early times, ii) {\it turn into a reduction of TDE rates for most of the relaxation time}. We also found that combining different PDMFs with either  ultra-steep stellar densities or radial velocity anisotropies only resulted in very minor changes.

%Hence, even top-heavy IMFs combined with either scenario cannot explain the post-starburst preference of TDEs. 

%with different PDMFs and found that enhancements and their evolution only slighcan increase by up to $\sim 10 \%$. 

%Hence we find that the scenarios invoking stellar properties: stellar velocity anisotropies \citep{Stone+18}, ultra-steep stellar densities \citep{Stone+18}, and even their combination with different PDMFs \footnote{The impact of different PDMFs was studied by \citep{Bortolas} albeit not in combination with other scenarios.} cannot reproduce the observed post-starburst preference of TDEs when taking into account strong scatterings. 

%In summary,
In summary, we have shown that stellar properties that were proposed to explain the post-starburst preference of TDEs, including stellar velocity anisotropies, ultra-steep stellar densities, and the combination of either with top-heavy PDMF \emph{cannot reproduce the observed enhancements, when taking into account both weak and strong scattering}. As we briefly discussed in Section \ref{sec:other} other explanations invoking the influence of a disk or a massive perturber also fail to reproduce either the strength or/and the duration of the enhancements observed (e.g., \cite{French20}). Hence, our work emphasizes both the importance of taking into account strong scatterings \citep{LC} and the need for new hypotheses to explain the post-starburst preference of TDEs.

\section*{Acknowledgements}
We would like to thank Aleksey Generozov for fruitful discussions. OT would like to thank Eugene Vasiliev for his support in the use of Phaseflow.

\bibliographystyle{mn2e}

\bibliography{lc_refs}

\appendix
%\onecolumn
\section{Local diffusion coefficient }
\label{app:diff}
The local diffusion coefficient we evaluate in Eq. \ref{eq:diffAvg} is given by \citep{MagorrianTremaine99, WangMerritt04}: 
\begin{equation}
\lim _{R \rightarrow 0} \frac{\left\langle(\Delta R)^{2}\right\rangle}{2 R}=\frac{32 \pi^{2} r^{2} G^{2}\left\langle m_{\star}^{2}\right\rangle \ln \Lambda}{3 J_{c}^{2}(\epsilon)}\left(3 I_{1 / 2}(\epsilon)-I_{3 / 2}(\epsilon)+2 I_{0}(\epsilon)\right)
\end{equation}
with: 
\begin{equation}
I_{0}(\epsilon) \equiv \int_{0}^{\epsilon} f\left(\epsilon^{\prime}\right) \mathrm{d} \epsilon
\end{equation}
\\
and 
\begin{equation}
I_{n / 2}(\epsilon) \equiv  {[2(\psi(r)-\epsilon)]^{-n / 2} } 
\int_{\epsilon}^{\psi(r)}\left[2\left(\psi(r)-\epsilon^{\prime}\right)\right]^{n / 2} f\left(\epsilon^{\prime}\right) \mathrm{d} \epsilon^{\prime}.
\end{equation}

\section{Orbit averaged ejection rates}
\label{app:analytics}
Here we present the the orbit averaged ejection rate for which we derived a simple analytical closed forms in \cite{LC}. 
 
For an equal mass scatterer and $ \gamma_\star = 3/2$, we obtained: 
 \begin{equation}
 \langle \dot{N}_{\rm ej} \rangle= \frac{2^{1/2} \pi \rho_{\rm infl} m G^{1/2} r_{\rm infl}^{3/2} (4 - 3 (1 - e^2)^{1/2})}{ (\gamma_\star+ 1)M ^{3/2}  (1-e^2)^{1/2}} .
\end{equation}
\\
For an equal mass scatterer, with $\gamma_\star= 5/2$, we obtained: 
 \begin{equation}
 \langle \dot{N}_{\rm ej} \rangle= \frac{2^{-1/2}\pi \rho_0 m G^{1/2} r_0^{5/2} (-4+ 12 e^2 + 5 (1-e^2)^{3/2})}{(\gamma_\star + 1) M ^{3/2} a (1-e^2)^{3/2}} .
\end{equation}

\section{Analytical solutions to the Fokker-Planck equation with strong scatterings}
\label{app:sol}
%\rev{We assume a slow variation with respect to $\tau$ such as df/d$\tau \approx$ 0,  and use the method of Frobenius to derive an analytic solution in angular momentum to the modified Fokker-Planck equation for these two power-law distributions.} 
In \citep{LC}, we derived analytical solutions for the modified Fokker-Planck equation Eq.\ref{eq:FP2} using the method of Frobenius. 
%and are valid at small  $j$, which is the relevant parameter space for stars that could undergo a disruption. 
A comparison between our analytical solutions and numerical solutions can be found in \citep{LC} (Fig.4 for equal mass scatterers,  Fig.5 for unequal mass scatterers).

\subsection{Equal mass scatterer}
%For the shallower slope, $\gamma_\star= 3/2$, the solution is very close to the Cohn-Kulsrud profile, whereas for the steeper slope of $\gamma_\star= 5/2$, it has an exponential behaviour. 
For equal mass scatterers, the analytical solutions for $\gamma_\star= 3/2$ and $\gamma_\star= 5/2$ write: 
\begin{equation}
\begin{split}
f_{3/2}(j) & = ( 1 + 16 A j) ( a + b \ln{j}) - 32 A b j 
\\
f_{5/2}(j) &=  e^{\frac{- 4 \sqrt{2 A}}{\sqrt{j}}} j^{1/4} [a g_{-}(j) + b e^{\frac{8 \sqrt{2 A}}{\sqrt{j}}} g_{+} (j) ]  
\label{eq:frobenius5.2}
 \end{split} 
\end{equation}
with  $g_{\pm} = 1 \pm \frac{\sqrt{j}}{32 \sqrt{ 2 A} } + 0.0022 \frac{j}{ A}$ and two undetermined constants -- $b$ and $a$ -- need to be found.  $A =  \langle \dot{N}_{ej} \rangle/ \mu (\epsilon)$ is the sink term whose analytical formula can be found in Appendix \ref{app:analytics}. 

Their explicit time-dependence comes from the undetermined constants $a$ and $b$.  In practice, $b$ is deterministically set as a function of $a$ using the absorbing boundary condition at $j=j_{o}$, so there is only one true time-dependent free parameter.  We find that $a$ depends on the dimensionless time $\tau$ as $a \sim \tau^{-1/2}$.

%Since the approximate time-scale for angular momentum relaxation to occur is $t_j(j)\sim j^2 t_{\rm r} $, our solution is accurate for $ j \lesssim j_{\rm CK} = \sqrt{\tau}$. In Fig. \ref{fig:analytical52equal}, the numerical solutions of the Fokker-Planck equation with strong scatterings (Eq. \ref{eq:FP2}) are compared to our analytical solution ($\gamma_\star=5/2$) at different dimensionless times. The agreement is excellent for values of $j \lesssim j_{\rm CK}$, the ``Cohn-Kulsrud'' angular momentum below which the distribution function has had time to relax into a QSS. 

\subsubsection{Unequal mass scatterer}
We derived analytical solutions to the modified Fokker-Planck equation Eq.\ref{eq:FP2} for the following stellar-mass black holes slopes: $\gamma_{\rm bh}=3/2 ,7/4 , 2, 9/4 $ and $5/2$. 
The resulting closed form solutions write: 
\begin{equation}
\begin{split}
f_{3/2,u}(j)&= ( 1 + 4 A j) (a + b \ln{j}) - 8 A j b 
\\
 f_{7/4,u}(j)&= (1 + 8 A \sqrt{j})^2 (a - \frac{b}{2}  (2 \ln{16 A} + \ln{j}))  
 \\
- & 2 b (\gamma_{E} + 16 A (-1 + \gamma_{E}) \sqrt{j} +  32 A^2 (-3 + 2 \gamma_{E}) j)  
 \\
f_{2,u}(j) & = j^{-2 \sqrt{A}} (j^{4  \sqrt{A}} a +b)
\\
f_{9/4,u} & = e^{-8 \sqrt{A} j^{-1/4}} j^{1/8} \left( \frac{ \sqrt{\pi b }} {2 A^{1/4}}  (-1 + \frac{ j^{1/4} }{ 64 A^{1/2}} ) \right.
\\
+ &  \left. \frac{a}{ 2 \sqrt{2 \pi}} \cosh{(8  \sqrt{A} j^{-1/4} ) } \right) 
\\
f_{5/2,u}(j)&=  e^{\frac{-4 \sqrt{A}}{\sqrt{j} }} j^{1/4} [a h_{-}(j) + b e^ { \frac{8 \sqrt{A}}{\sqrt{j} }} h_{-} (j)]
\end{split}
\label{eq:frobeniusunequal}
\end{equation}
where $\gamma_E$ is Euler's constant and $h_{\pm} = 1 \pm \frac{\sqrt{j}}{32 \sqrt{A} } + 0.0044 \frac{j}{ A} $.
\end{document}